\documentclass[11pt,a4paper]{article}
\usepackage{cfm}
\begin{document}
\cfmtitre{
 Dynamical pressure for fluid mixtures\\ with several temperatures}

\cfmauteur{Henri Gouin$^{a}$, Tommaso Ruggeri$^{b}$}
\cfmadresse{$a.$Aix-Marseille Universit\'e, CNRS, Centrale Marseille, M2P2 UMR 7340,\ \ 13451  Marseille, France\\
  Email: {henri.gouin@univ-amu.fr; henri.gouin@yahoo.fr}\\
$b.$ Department of Mathematics and Research Center of Applied
Mathematics C.I.R.A.M. University\, of Bologna, Via Saragozza 8,
40123-I Bologna\\
Email: {tommaso.ruggeri@unibo.it}}

\cfmabstract{We consider binary mixtures of fluids   with
  components having different temperatures. A new dynamical pressure term is associated with the
difference of temperatures between components even if fluid
viscosities are null. The non-equilibrium dynamical pressure can be
measured and   may be convenient in several physical situations   as
for example in cosmological circumstances where  a dynamical
pressure played a major role in the evolution of the early
universe.}\\

{\textbf{Key words}: \emph{Fluid mixtures;
Multi-temperatures; Dynamical pressure; Hamilton's principle.}}\\
\cfmresume{Nous considérons des mélanges de fluides où chaque
constituant a sa propre température. La différence de températures
entre constituants implique l'existence d'une nouvelle pression
dynamique même si les fluides ont une viscosité nulle. Cette
pression dynamique peut être mesurée et utile dans de nombreuses
situations physiques comme en cosmologie où une pression dynamique
joue un rôle majeur dans l'évolution des débuts de l'univers. }
\\

{\textbf{Mots clés}: \emph{Mélanges de fluides; multi-temperatures; pression dynamique; principe d'Hamilton.}}

\section{Introduction.}

The theory of mixtures  generally considers two different kinds of
continua: homogeneous mixtures (each component occupies the whole
mixture volume) and heterogeneous ones (each component occupies only
a part of the mixture volume). At least four approaches to the
construction of two-fluids models are known.\newline The first one
for studying the heterogeneous two-phase flows is an averaging method
\cite{Ishii}.   A second approach  was used for the construction of
a quantum liquid model and was purposed for the homogeneous mixtures
of fluids \cite{Putterman}.    A third approach is   done in the
context of rational thermodynamics founded on the postulate that
each constituent obeys the same balance laws as a single fluid \cite
{ET,RS}.

At least, there exists a different approach based on the Hamilton principle
which is used for the construction of conservative  mathematical
models of continua.  The variations of the Hamilton action are
constructed in terms of virtual motions of continua which may be
defined both in Lagrangian and Eulerian coordinates
\cite{Serrin}.\newline Here, we use variations in the case of fluid
mixtures. The variational approach to the construction of two-fluid
models has been used by many
authors    \cite%
{Berdichevsky,gouin2,Gouin-Ruggeri}. The method is different from the
method proposed is \cite{GR} and is now developed in \cite{GR1}.
\newline To study thermodynamical processes by using the Hamilton principle,
 the entropies of components are
added to the field parameters instead of temperatures. The
Lagrangian is the difference between the kinetic energy and an
internal potential per unit volume depending on the densities, the
entropies and the relative velocities of the mixture components. It
is not necessary to distinguish
 molecular mixtures from heterogeneous fluids when each component
occupies only a part of the mixture volume \cite{Gavrilyuk 2}. The
terms including interaction between different components  come from
the direct knowledge of the internal potential.
\newline The assumption of a common temperature for all the
components is open to doubt for the suspensions of particles
\cite{lhuillier} as well as in the mixtures of gases in the early
universe \cite{Groot}. By using the Hamilton principle, the
existence of several temperatures (one temperature for each
component) must be associated with the existence of several
entropies (one specific entropy for each component). The internal
potential per unit volume is a function of the densities, the
entropies and the difference of velocity between components.

\section{Governing equations in conservative cases.}

In a Galilean system of coordinates, the motion of a two-fluid
mixture is represented as
$$
\mathbf{Z}_{\alpha }=\mathbf{\Phi }_{\alpha
}(\mathbf{z})\,,\qquad\quad (\alpha=1, 2)  \label{representation}
$$
where $\mathbf{z}=(t,\mathbf{x)}$ denotes Eulerian coordinates in a
four-dimensional domain $\omega$ in the time-space and $\mathbf{Z}%
_\alpha=(\lambda,\mathbf{X}_\alpha)$ denotes Lagrangian coordinates
of the component $\alpha$ in a four-dimensional reference space
$\omega_\alpha$. The conservation of matter for each component
requires
\begin{equation}
\rho _{\alpha }\,\rm{ det}\,F_a =\rho _{\alpha 0}\left(
\mathbf{X}_{\alpha
}\right)\quad \mathrm{with}\quad F_a= \frac{\partial \mathbf{x}}{%
\partial \mathbf{X}_{\alpha }},  \label{mass0}
\end{equation}
where index $_{\alpha 0} $ corresponds to the reference density in
$\omega_{\alpha }$ and
$\rm{ det}\left( {\partial \mathbf{x}}/{\partial \mathbf{X}_{\alpha }}%
\right) $ is the Jacobian determinant of the motion of the component
$\alpha$ of density $\rho _{\alpha }$. In differentiable cases Eqs
(\ref{mass0}) are equivalent to the equations of density balances
\begin{equation}
{\frac{\partial \rho _{\alpha }}{\partial t}}\ +\ \rm{div}(\rho _{\alpha }{%
\mathbf{v}}_{\alpha })=0 ,  \label{mass i}
\end{equation}
where ${\mathbf{v}}_{\alpha }$ denotes the velocity of each component $%
\alpha $. The Lagrangian of the binary system is
$$
L = \sum_{\alpha =1}^{2}\, \left(\frac{1}{2} \ \rho _{\alpha }\,{\mathbf{v}}%
_{\alpha }^{2}-\rho _{\alpha }\Omega _{\alpha }\right) -\eta (\rho
_{1},\rho _{2},s_{1},s_{2},{\mathbf{u}}),
$$
where the summation is taken over the fluid components $(\alpha
=1,\,2)$,
$\,s_{\alpha }$ are the specific entropies, $\,{\mathbf{u}} = {\mathbf{v}}%
_{2}-{\mathbf{v}}_{1}\ $ is the relative velocity of components,
$\,\Omega _{\alpha }\ $ are the external force potentials, $\,\eta $
is a potential
per unit volume of the mixture. The Lagrangian $\,L\,$ is a function of $%
\rho _{\alpha },{\mathbf{v}}_{\alpha },s_{\alpha }$ and we introduce
the quantities
\begin{equation}
R_{\alpha }  \equiv  \displaystyle\frac{\partial L}{\partial \rho
_{\alpha }}  =\frac{1}{2}\, {\mathbf{v}}_{\alpha }^{2}-
\frac{\partial \eta }{\partial \rho _{\alpha }}  -  \Omega _{\alpha
},  \label{R} \quad
{\mathbf{k}}_{\alpha }^{T}\ \equiv \ \displaystyle\frac{1}{\rho _{\alpha }}%
\ \frac{\partial L}{\partial {\mathbf{v}}_{\alpha }}  =  {\mathbf{v}}%
_{\alpha }^{T}  -\frac{(-1)^{\alpha }}{\rho _{\alpha }}
\frac{\partial \eta }{\partial {\mathbf{u}}},  \quad
 \rho _{\alpha }\, T_{\alpha }  \equiv   \displaystyle-\frac{\partial L}{%
\partial s_{\alpha }}  = \frac{\partial \eta }{\partial s_{\alpha }},
\label{T}
\end{equation}
where $^T$ denotes the transposition and $\ \displaystyle
{\partial L}/{\partial {\mathbf{v}}_{\alpha }},\  {\partial \eta }/{%
\partial {\mathbf{u}}}\ $ are linear forms.  Relation (\ref{T})$_3$ defines the
temperatures $T_{\alpha } \ (\alpha=1,2)$ which are dynamical
quantities depending on $\rho_1,\rho_2,s_1,s_2$ and
${\mathbf{u}}$.

\medskip
To obtain the equations of component motions by means of the
Hamilton principle, we consider variations of particle motions in
the form of surjective mappings  $\, \mathbf{X}_{\alpha
}=\mathbf{\Xi }_{\alpha }(t,\mathbf{x;} \kappa _{\alpha }), $ where
scalars $\kappa _{\alpha }$ are defined in a neighborhood of zero;
they are associated with a two-parameter family of virtual motions.
The real
motions correspond to $\kappa _{\alpha }=0\, $ such that $\  \mathbf{\Xi }_{\alpha }(t,%
\mathbf{x;}0)=\phi _{\alpha }(t,\mathbf{x)}\ $ and virtual Lagrange
displacements are
\begin{equation}
{\delta _{\alpha }}\mathbf{X}_{\alpha }=\frac{\partial \mathbf{\Xi
}_{\alpha
}(t,\mathbf{x;}\kappa _{\alpha })}{\partial \kappa _{\alpha }}%
\left\vert _{\kappa _{\alpha }=0}\right., \quad (\alpha=1, 2).
\label{displacement1}
\end{equation}%
The Hamilton action is $\ \displaystyle a=\int_{\omega }\ L\
dv\,dt.\ $ \emph{We first consider the Hamilton principle in the
form}
$$
{\delta _{\alpha }}a\equiv \left( \frac{da}{d\kappa _{\alpha
}}\right) _{|_{\kappa _{\alpha }=0}}=\ {\delta _{\alpha
}}\int_{\omega }\ L\ dv\,dt\ =\ 0,
$$
under constraints (\ref{mass0}),  where ${\delta _{\alpha }}a\,$ are
the variations of $a$ associated with Rel. (\ref{displacement1}).
\newline From the definition of virtual motions, we obtain
the values of $\ {\delta _{\alpha }}{\mathbf{v}}_{\alpha
}\,({\mathbf{x}},t)$, ${\delta
_{\alpha }}\rho _{\alpha }\,({\mathbf{x}},t)$ and ${\delta _{\alpha }}%
s_{\alpha }\,({\mathbf{x}},t)\,\;$where the notation ${\delta _{\alpha }}b(t,{\mathbf{x}}%
) $ represents the variation of function $b$ at $(t,\,{\mathbf{x}})$
fixed.
  The functions are assumed to be
smooth enough in the domain $\,\omega _{\alpha }\,$ and $\ {\delta
_{\alpha }}{\mathbf{X}}_{\alpha }\ =\ 0\ $ on its boundary.  It
follows \cite{GR1},
$$
{\delta _{\alpha }}a=\int_{\omega _{\alpha }}  \rho _{\alpha 0}\left( -%
\frac{\partial R_{\alpha }}{\partial {\mathbf{X}}_{\alpha
}}+\frac{\partial
}{\partial \lambda }({\mathbf{k}}_{\alpha }^{T}\ F_{\alpha })-T_{\alpha }%
\frac{\partial s_{\alpha 0}}{\partial {\mathbf{X}}_{\alpha }}\right)
{\delta _{\alpha }}{\mathbf{X}}_{\alpha }  \ dv_{\alpha }\,dt,
$$%
and we get
the component motion equations  in Lagrangian coordinates,
$\displaystyle
\frac{\partial }{\partial \lambda }({\mathbf{k}}_{\alpha }^{T}\ F_{\alpha })-%
\frac{\partial R_{\alpha }}{\partial {\mathbf{X}}_{\alpha }}-T_{\alpha }\,%
\frac{\partial s_{\alpha 0}}{\partial {\mathbf{X}}_{\alpha }}\ =0.\
$
 By
taking into account the identity $\displaystyle\ {\frac{d_{\alpha
}F_{\alpha }}{dt}}\,-\,{\frac{\partial {\mathbf{v}}_{\alpha
}}{\partial {\mathbf{x}}}}\ F_{\alpha }=0$ and for $\lambda =t$, we
rewrite the equations in Eulerian coordinates,
$$
\frac{d_{\alpha }{\mathbf{k}}_{\alpha }^{T}}{dt}\
+\,{\mathbf{k}}_{\alpha
}^{T}\ \frac{\partial {\mathbf{v}}_{\alpha }}{\partial {\mathbf{x}}}\,=\,%
\frac{\partial R_{\alpha }}{\partial {\mathbf{x}}}\ +T_{\alpha }\frac{%
\partial s_{\alpha }}{\partial {\mathbf{x}}}\ .  \label{(motions)}
$$
To obtain the equation of energy, \emph{we need a second variation
of motions associated with the time parameter}. The variation
corresponds to a virtual motion in the form $\ \lambda =\varphi
(t;\kappa ), $ where scalar $\kappa $ is defined in a neighborhood
of zero. The real motion of the mixture corresponds to $\kappa =0$
such that $\varphi (t,0)=t$; the associated virtual displacement is
$$
\delta \lambda =\frac{\partial \varphi (t;\kappa )}{\partial \kappa }%
\left\vert _{\kappa =0}\right. .
$$%
 In multi-component fluids, due to exchanges of
energy between the components, the entropies cannot be conserved
along component paths; in the reference spaces $\omega _{\alpha }$,
the specific entropies $s_{\alpha }$ depend also on $\lambda $
$$
s_{\alpha }=s_{\alpha 0}\left( \lambda ,\mathbf{X}_{\alpha }\right)
.
$$%
\emph{The variation of  Hamilton's action associated with the second
family of virtual motions} yields
$$
{\delta }a\equiv \ {\delta }\int_{\omega }\ L\ dv\,dt\ =\
\int_{\omega }\   \frac{\partial L}{\partial \lambda }\ \delta
\lambda   \ dv\,dt\,=0 .
$$%
From $\displaystyle  \frac{\partial L}{\partial \lambda
}=\sum_{\alpha=1}^{2}  \frac{\partial L}{\partial s_{\alpha }}\, \frac{\partial s_{\alpha 0}%
}{\partial \lambda }$, we deduce when $\lambda=t$, $\displaystyle  \frac{\partial L}{\partial \lambda }= -\sum_{\alpha =1}^{2}\ \rho _{\alpha }T_{\alpha }\frac{%
d_{\alpha }s_{\alpha }}{dt}$,   where\ $\displaystyle
\frac{d_{\alpha }s_{\alpha }}{dt}=\frac{\partial s_{\alpha
}}{\partial t}+\frac{\partial s_{\alpha }}{\partial
\mathbf{x}}\,\mathbf{v}_{\alpha }$  is the material derivative with
respect to velocity
 $\mathbf{v}_{\alpha }$.
 We obtain  for the
total mixture
\begin{equation}
\sum_{\alpha =1}^{2} \rho _{\alpha }T_{\alpha }\frac{d_{\alpha }s_{\alpha }}{%
dt} =0 .  \label{total energy01}
\end{equation}
Due to Eqs. (\ref{mass i}) we obtain the equivalent form
\begin{equation}
\sum_{\alpha =1}^{2} Q_\alpha=0\qquad \mathrm{with} \qquad
Q_\alpha=\left( \frac{\partial \rho _{\alpha }s_{\alpha }}{\partial
t}+\mathrm{{div}(\rho _{\alpha }s_{\alpha }\mathbf{v}_{\alpha
})}\right) \,T_{\alpha } . \label{total energy1}
\end{equation}%
Equation (\ref{total energy1}) expresses that the exchange of energy
between components has a null total amount. In case of mixtures with
two entropies,  the Hamilton principle is not able to close the
system of motion equations;
we need additional arguments to obtain the evolution equations for each entropy $%
s_{\alpha }$ by considering the behaviors of $Q_{\alpha }$.  A
possibility to close the system of equations is to consider that the
momenta and heat exchanges between the   components are rapid enough
to have a common temperature.   Another possibility, used by Landau
for quantum fluids \cite{Putterman}, is to assume that the total
specific entropy $s$ is convected along the first component
trajectory.  These assumptions are not valid for heterogeneous
mixtures where each phase has different pressures and temperatures
\cite{lhuillier,Groot}.

\section{Mixtures weakly out of equilibrium.}
We consider  the case when the mixture is \emph{weakly out of
equilibrium} such that the difference of velocities $\mathbf{u}$ and
the difference of temperatures $T_{2}-T_{1}$ are small enough with
respect to the main field variables.\\
 In the following,
$\displaystyle \rho\, {\mathbf{v}}= \sum_{\alpha =1}^{2}\rho
_{\alpha }{\mathbf{k}}_{\alpha }=\sum_{\alpha =1}^{2}\rho _{\alpha
}{\mathbf{v}}_{\alpha }$ is the total momentum and
$\displaystyle\rho =\sum_{\alpha =1}^{2}\rho _{\alpha }\ $ is the
 mixture density.

   For the sake of simplicity, we  neglect the external
forces. Generally, the volume potential $\eta $ is developed in the
form  \footnote{In Ref. \cite{GR}, the internal energy is the sum of
the internal energies of the components
 $\left(\rho\,\varepsilon = \sum_{\alpha=1}^2
\rho_\alpha\varepsilon_\alpha(\rho_\alpha,s_\alpha)\right)$.}
$$
\eta (\rho _{1},\rho _{2},s_{1},s_{2},\mathbf{u})=e(\rho _{1},\rho
_{2},s_{1},s_{2})-b(\rho _{1},\rho _{2},s_{1},s_{2})\,\mathbf{u}^{2}
,
$$%
where $b$ is a positive function of $\rho _{1},\rho
_{2},s_{1},s_{2}$.
 We consider the linear
approximation when
$|\mathbf{u}|$ is small with respect to $|\mathbf{v}_{1}|$ and $|\mathbf{v}%
_{2}|$. In linear approximation the volume potential is  equal to
the volume internal energy $e$ \cite{Gavrilyuk 2},
$$
\eta (\rho _{1},\rho _{2},s_{1},s_{2},\mathbf{u}) \approx e(\rho
_{1},\rho _{2},s_{1},s_{2})=\rho\,\varepsilon(\rho _{1},\rho
_{2},s_{1},s_{2}),
$$
where $\varepsilon$ denotes the internal energy per unit mass. Let
us note that the diffusion vector $\mathbf{j}=\rho
_{1}(\mathbf{v}_{1}-\mathbf{v})\equiv \rho
_{2}(\mathbf{v}-\mathbf{v}_{2})$ is a small momentum vector deduced
both from velocities and densities of the components. The
equations of density balances (\ref{mass0}) can be written in the
form
\begin{equation}
\frac{d\rho }{dt}+\rho \,\mathrm{div}\,\mathbf{v}=0 \qquad \mathrm{and}%
\qquad \rho \,{\frac{dc}{dt}}+\mathrm{div}\,{\mathbf{j}}=0 ,
\label{masses}
\end{equation}%
\newline
where $\displaystyle\, c= {\rho _{1}}/{\rho} $ denotes the
concentration of component $1$ and $\displaystyle  {\emph{d}
}/{\emph{dt}}={\partial}/{\partial \emph{t}}+ {\partial}/{\partial
\mathbf{x}}\,.\,{\mathbf{v}}$ is the material derivative with
respect to the average velocity of the mixture.
The divergence of a linear operator $\mathbf{A}$ is the covector $\mathrm{div%
} \mathbf{A }$ such that, for any constant vector ${\mathbf{a}}, $ $(\mathrm{%
div}\, \mathbf{A})\, {\mathbf{a}} = \mathrm{div }\, (\mathbf{A}\ {\mathbf{a}}%
)$ and we write ${\mathbf{v}}_{\alpha }{\mathbf{v}}_{\alpha }^T \equiv{%
\mathbf{v}}_{\alpha }\otimes {\mathbf{v}}_{\alpha }$.   Let us
denote by $\displaystyle h_{\alpha }\equiv  {\partial e}/{\partial
\rho _{\alpha }}\ $ the specific enthalpy of the component $\alpha$.
 For processes with weak diffusion, the equations of
component motions get the form,
$$
\rho _{\alpha }\,\mathbf{\Gamma }_{\alpha }\equiv \frac{\partial {\rho }_{{%
\alpha }}{\mathbf{v}}_{\alpha }}{\partial t}\,+\,\mathrm{div}(\rho _{\alpha }%
{\mathbf{v}}_{\alpha }\otimes {\mathbf{v}}_{\alpha })^{T}=\rho
_{\alpha
}T_{\alpha }\,\mathrm{{grad}\,\emph{s}_{\alpha }-\rho _{\alpha }\,grad\,%
\emph{h}_{\alpha } }.  \label{component equation of motions i}
$$%
The equation of total momentum  is \cite{GR1}:  $\qquad
\displaystyle \frac{\partial {\rho \mathbf{v}}}{\partial
t}\,+\,\mathrm{div}\left(
\sum_{\alpha =1}^{2}\left( \rho_\alpha   {\mathbf{v}}_{\alpha }\otimes {\mathbf{v}}%
_{\alpha }\right) -\mathbf{t}\right) ^{T}=0, $\\
where $\ \mathbf{t}=\sum_{\alpha =1}^{2}%
\mathbf{t}_{\alpha }\ $ is the total stress tensor such that $
\mathbf{t}_{\alpha \nu \gamma }=-p_{\alpha }\;\delta _{\nu \gamma
},\,
  p_{\alpha }=
\rho\rho_\alpha\varepsilon_{,\rho_\alpha}=\rho _{\alpha }e_{,\rho
_{\alpha }}-{\rho_{\alpha }e}/{\rho }\,,\  p=\sum_{\alpha
=1}^{2}p_{\alpha }\, .  $ The internal energy is a natural function
of densities and entropies. Due to Def. (\ref{T})$_3$,
\begin{equation}
\rho_1\,T_{1}= \rho\, \frac{\partial \varepsilon}{\partial
s_{1}}(\rho _{1},\rho _{2},s_1,s_2)\quad \mathrm{and}\quad
\rho_2\,T_{2}= \rho\, \frac{\partial \varepsilon}{\partial
s_{2}}(\rho _{1},\rho _{2},s_1,s_2).\label{Temperatures}
\end{equation}
Let us denote by $\overline{\varepsilon}$ the expression of the
specific internal energy as a function of $\rho, c, s_1, s_2$ such
that $\overline{\varepsilon}(\rho,c,s_1,s_2)=
 \varepsilon(\rho_1,\rho_2,s_1,s_2)$; we get:
$$
\rho \,\frac{d\varepsilon }{dt}= \rho\, \frac{\partial
\overline{\varepsilon}}{\partial \rho}\,\frac{d\rho }{dt}+\rho\,
\frac{\partial \overline{\varepsilon}}{\partial c}\,\frac{dc
}{dt}+\rho\, \frac{\partial \overline{\varepsilon}}{\partial
s_1}\,\frac{ds_1}{dt}+\rho\, \frac{\partial
\overline{\varepsilon}}{\partial s_2}\,\frac{ds_2}{dt}.
\label{Gibbsa}
$$
Due to the fact that $\displaystyle \rho^2\, \frac{\partial
\overline{\varepsilon}}{\partial \rho} = p\ $ and $\ \displaystyle
\frac{\partial \overline{\varepsilon}}{\partial c}= h_1-h_2$, we
obtain
\begin{equation}
\rho \,\frac{d\varepsilon }{dt}= \frac{p}{\rho}\,\frac{d\rho
}{dt}+\rho\, (h_1-h_2)\,\frac{dc }{dt}+\rho_1\, T_1
\,\frac{ds_1}{dt}+\rho_2\, T_2\,\frac{ds_2}{dt}. \label{Gibbsb}
\end{equation}
By taking into account that $\ \displaystyle \frac{d_\alpha
s_\alpha}{dt}=\frac{ds_\alpha}{dt}+ \frac{\partial
s_\alpha}{\partial \mathbf{x}}(\mathbf{v}_\alpha-\mathbf{v})\ $ and
by using Eqs. (\ref{total energy01}), (\ref{masses}), Eq.
(\ref{Gibbsb}) yields
\begin{equation}
\rho \,\frac{d\varepsilon }{dt}+p\
\mathrm{div}\,\mathbf{v}\,+(h_{1}-h_{2})\
\mathrm{div}\,\mathbf{j}+(T_{1}\,\mathrm{{grad}\,\emph{s}_{1}-T_{2}\,{grad}\,%
\emph{s}_{2})^{\emph{T}}\,\mathbf{j}=0 }.  \label{Total energy4}
\end{equation}

Due to Eqs. (\ref{Temperatures}), the internal energy can be
expressed as a function of densities and temperatures of components
$$
\tilde{\varepsilon} (\rho _{1},\rho _{2},T_1,T_2) = \varepsilon
(\rho _{1},\rho _{2},s_1,s_2).
$$
As we wrote in  \cite{GR}, we define the average temperature $T$
associated with $T_1$ and $T_2$ through the implicit solution of the
equation
\begin{equation}
\tilde{\varepsilon} (\rho _{1},\rho _{2},T ,T)=\tilde{\varepsilon}
(\rho _{1},\rho _{2},T_1,T_2).  \label{averageT}
\end{equation}
We denote by $\Theta_\alpha= T_\alpha-T$ the difference between
component and average temperatures, which are non-equilibrium
thermodynamical variables. Near equilibrium, Eq. (\ref{averageT})
can be expanded to the first order; then
\begin{equation}
\sum_{\alpha =1}^{2} c_v^\alpha \, \Theta_\alpha = 0\qquad
 {\rm with}\qquad c_v^\alpha  =
 \frac{\partial \widetilde{\varepsilon}}{\partial T_\alpha}(\rho _{1},\rho _{2},T ,T).\label{linearized}
\end{equation}
Due to the fact that $\ \displaystyle
 \rho\, d\varepsilon =\sum_{\alpha
 =1}^{2}\rho_\alpha\,T_\alpha\,ds_\alpha +
 \frac{p_\alpha}{\rho_\alpha}\, d\rho_\alpha,\
$ then
\begin{equation}
\rho\, c_v^1  = T \sum_{\alpha =1}^{2}\rho_\alpha\,
 \frac{\partial  {s_\alpha}}{\partial T_1}(\rho _{1},\rho _{2},T
 ,T)\ \qquad
 {\rm and}\qquad \ \rho\, c_v^2  = T\sum_{\alpha =1}^{2}\rho_\alpha\,
 \frac{\partial  {s_\alpha}}{\partial T_2}(\rho _{1},\rho _{2},T
 ,T).
 \label{c}
\end{equation}
The definition of the total entropy $s$ of the mixture is
\begin{equation}
\rho\,s =\sum_{\alpha =1}^{2}\rho_\alpha s_\alpha(\rho _{1},\rho
_{2},T_1,T_2).\label{average entropy3}
\end{equation}
The first order expansion of Eq. (\ref{average entropy3}) yields
$$
\rho\,s =\sum_{\alpha =1}^{2}\rho_\alpha s_\alpha(\rho _{1},\rho
_{2},T,T)+ \rho_\alpha s_\alpha \frac{\partial  {s_\alpha}}{\partial
T_1}(\rho _{1},\rho _{2},T ,T)\, \Theta_1+ \rho_\alpha s_\alpha
\frac{\partial  {s_\alpha}}{\partial T_2}(\rho _{1},\rho _{2},T
,T)\, \Theta_2.\label{average entropy4}
$$
Due to Rels. (\ref{linearized}), (\ref{c}), $\ \displaystyle \rho\,s
=\sum_{\alpha =1}^{2}\rho_\alpha s_\alpha(\rho _{1},\rho _{2},T,T)\
$ and the specific entropy $s$ does not depend on $\Theta_1$ and
$\Theta_2$ but only on $\rho_1, \rho_2$ and $T$. We denote by
$\hat{{\varepsilon}}$ the internal specific energy as a function of
$\rho, c, T$:
$$
\hat{{\varepsilon}} (\rho,c,T)= \tilde{\varepsilon} (\rho _{1},\rho
_{2},T ,T),
$$
which satisfies the Gibbs equation
$$
Tds=d\hat{\varepsilon}-\frac{p_{o}}{\rho ^{2}}d\rho +\left( \mu
_{2}-\mu _{1}\right) dc
$$
where $p_{o}\left( \rho ,c,T\right) $ is the equilibrium pressure at
temperature $T$ and $\mu _{2}-\mu _{1}$, difference of component
chemical potentials,  is the chemical potential of the whole
mixture. By taking into account of equation (\ref{masses}), we get
$$
\rho \,\frac{d\hat{\varepsilon}}{dt}+p_{o}\
\mathrm{div}\,\mathbf{v}+\left(
\mu _{1}-\mu _{2}\right) \ \mathrm{div}\,\mathbf{j}-\rho \,T\,\frac{ds}{dt}%
=0.
$$%
Moreover,
\begin{equation}
\rho \,\frac{ds}{dt}=\sum_{\alpha =1}^{2}\rho _{\alpha }
\,\frac{d_{\alpha }s_{\alpha
}}{dt}+\mathrm{div}\,[(s_{2}-s_{1})\mathbf{j}\,] .\label{entropies}
\end{equation}

Equation (\ref{entropies}) yields the relation between the material
derivatives of entropy $s$ and entropies $s_1$ and $s_2$. By taking
into account   this result  in Eq. (\ref{Total energy4}) and
$\hat{\varepsilon}(\rho,c,T) = \varepsilon(\rho_1,\rho_2,s_1,s_2)$,
we obtain
\begin{equation}
T \sum_{\alpha =1}^{2}\rho _{\alpha } \frac{d_{\alpha }s_{\alpha }}{dt}%
+\left( p-p_{o}\right) \ \mathrm{div}\,\mathbf{v}\,+\Big(
(h_{1}-h_{2})-\left( \mu _{1}-\mu _{2}\right) +T\,(s_{2}-s_{1})\Big) \,%
\mathrm{div}\,\mathbf{j}
+\big(\Theta_1 \mathrm{{grad}\emph{s}_{1}-\Theta_2{grad}%
 \emph{s}_{2}\big)^{T}\mathbf{j}=0}.  \label{Gibbs1}
\end{equation}
The differences of temperatures $\Theta_1\equiv T_{1}-T$ and
$\Theta_2\equiv T_{2}-T$ are small with respect to $T$ and
$\mathbf{j}$\, is a small diffusion term with respect to the mixture
momentum $\rho\mathbf{v}$; consequently, in an approximation to the
first order, the term
$$
\big(\Theta_1\,\mathrm{{grad}\,\emph{s}_{1}-\Theta_2\,{grad}\,\emph{s}_{2}%
\big)^{\emph{T}}\ \mathbf{j}\ \,}
$$
is negligible. Let us consider
$$
K \equiv \Big((h_{1}-h_{2})-(\mu_1-\mu_2)%
+T\,(s_{2}-s_{1})\Big)\,\mathrm{div}\,\mathbf{j}\ ;
$$
we get
$$
K = \Big(\big(h_{1}-T_1s_1\big)
-\big(h_{2}-T_2s_2\big)-(\mu_1-\mu_2)+\Theta_1 s_1+\Theta_2
s_2\Big)\,\mathrm{div}\,\mathbf{j}\ .
$$
In an approximation to the first order, the term $ \big(\Theta_1
s_1+\Theta_2 s_2\big)\,\mathrm{div}\,\mathbf{j}\ $
 is
negligible.\\

 Due to the fact that $%
\mu_\alpha (\rho_1,\rho_2,T_1,T_2) = h_\alpha-T_\alpha s_\alpha$ is
the chemical potential of the component $\alpha$,   when
$\mathbf{j}$\, is a small diffusion velocity with respect to average
velocity $\mathbf{v} $, the term
$$
\displaystyle\bigg(%
 \mu_{1}(\rho_1,\rho_2,T_1,T_2)-\mu_{2}(\rho_1,\rho_2,T_1,T_2)
 -\Big(\mu_1(\rho_1,\rho_2,T) -\mu_2(\rho_1,\rho_2,T)\Big)\bigg)%
\,\mathrm{div}\,\mathbf{j}
$$
 is vanishing in an approximation to the
first order.

Consequently,  in an approximation to the first order, Eq. (\ref%
{Gibbs1}) reduces to
\begin{equation}
\sum_{\alpha =1}^{2}\rho _{\alpha }\,\frac{d_{\alpha }s_{\alpha }}{dt}=-%
\frac{1}{T}\left( p-p_o%
\right) \ \mathrm{div}\,\mathbf{v} . \label{Planck}
\end{equation}
The exchange of energy between components must obey   the second law
of thermodynamics: the total entropy rate is an increasing function
of time and we consider the second law of thermodynamics in the form
\begin{equation}
\sum_{\alpha =1}^{2}\left( \frac{\partial \rho _{\alpha }s_{\alpha }}{%
\partial t}+\mathrm{{div}(\rho _{\alpha }s_{\alpha }\mathbf{v}_{\alpha })}%
\right) \,\geq 0  \label{CD1}
\end{equation}
Due to Rels. (\ref{mass i}) the Clausius-Duhem inequality
(\ref{CD1}) is equivalent to
$$
\sum_{\alpha =1}^{2}\rho _{\alpha }\,\frac{d_{\alpha }s_{\alpha
}}{dt}\,\geq 0 \,. \label{CD2}
$$
This implies that the second member of Rel. (\ref{Planck}) must be
positive. Therefore, as usual in thermodynamics of irreversible
processes, the entropy inequality requires
\begin{equation}
\pi \equiv p-p_o  =-\Lambda \ \mathrm{div}\,\mathbf{v}\, .
\label{other pressure}
\end{equation}%
This expression defines the Lagrange multiplier $\Lambda $ of
proportionality such that $\Lambda \geq 0$. The dynamical pressure
$\pi $ is the difference between the pressure in the process out of
equilibrium with different temperatures for the components and the
pressure of the mixture assumed in local thermodynamical equilibrium
with the common average temperature $T$. Let us notice that Eqs.
(\ref{total energy1},\ref{other pressure})   allow to obtain
$Q_{\alpha }$ values. In fact,
$$
\rho _{1}T\,(T_{2}-T_{1})\frac{d_{1}\emph{s}_{1}}{dt}=\Lambda \,T_{2}\ (%
\mathrm{div}\,\mathbf{v)}^{2}\quad  {\rm{and}}\mathrm{\quad \rho _{2}%
\emph{T}\,(\emph{T}_{1}-\emph{T}_{2})\frac{d_{2}\emph{s}_{2}}{dt}=\Lambda \,%
\emph{T}_{1}\ (\mathrm{div}\,\mathbf{v)}^{2}}
$$%
and the system of field equations is now closed.

\section{Conclusion.}

The Hamilton principle points out that a dynamical pressure can be
obtained by neglecting viscosity, friction or external heat fluxes.
This is a main property of mixtures with multi-temperatures and this
fact may have some applications in plasma of gases and in the
evolution of the early universe \cite{Weinberg}.
\\
The results are in complete accordance with the ones  by Gouin \&
Ruggeri \cite{GR}  and developed in \cite{GR1}. This is an important
verification of the fact that the Hamilton principle can be extended
to nonconservative mixture motions when components have different
temperature. A difference with classical thermodynamics methods
 is that the volume internal energy   is not necessary the
sum of the volume internal energies of the components. In
\cite{GR1}, the volume internal energy is a nonseparate function of
densities and entropies (or temperatures) and   is consequently more
general than in   \cite{GR}.
\newline

\bigskip

\end{document}